%Title: Fusion Rings Related to Affine Weyl Groups 
%Authors: P. Furlan and V.B. Petkova
%Comments: 13 pages; contains three files: TeX file and two .eps files; 
%TeX file uses as input files: harvmac.tex, epsf.tex, amssym.def, 
%amssym.tex   
%(Talk  at ``Lie Theory and Its Applications in Physics III'',
%Clausthal,  11-14 July, 1999, to appear in the Proceedings,  eds.
%H.-D. Doebner et al)

\input harvmac 

%%%%%%%%%%%%%%%%%%%%%%%%%%%%%%%%%%%%%%%%%%%%%%%%%%%%%%%%%%%%%%%%
\input epsf.tex
%%%%%%%%%%%%%%%%%%%%%% macros for figures %%%%%%%%%%%%%%%%%%%%%%%
\newcount\figno
\figno=0
\def\fig#1#2#3{
\par\begingroup\parindent=0pt\leftskip=1cm\rightskip=1cm\parindent=0pt
\baselineskip=11pt
\global\advance\figno by 1
\midinsert
\epsfxsize=#3
\centerline{\epsfbox{#2}}
\vskip 12pt
\centerline{{\bf Fig. \the\figno}} 
#1\par
\endinsert\endgroup\par}
\def\figlabel#1{\xdef#1{\the\figno}}
\def\encadremath#1{\vbox{\hrule\hbox{\vrule\kern8pt\vbox{\kern8pt
\hbox{$\displaystyle #1$}\kern8pt}
\kern8pt\vrule}\hrule}}

%%%%%%%%%%%%%%%%%%%DEFINITIONS%%%%%%%%%%%%%%%%%%%%%%%%%%%%%%%%%

\def\blank#1{}
\def\za{\alpha}  
 \def\lla{\overline{\Lambda}}

%%%%%%%%%%%%%%%%%%%%%
\def\za{\alpha}  \def\zg{\gamma} 
   
\def\zk{\kappa} \def\zl{\lambda}  
  \def\zr{\rho}

\def\bL{\bar{\Lambda}}
\def\zG{\Gamma}   
\def\zL{\Lambda}  

\def\IZ{Z\!\!\!Z}
\def\[{\,[\!\!\![\,} \def\]{\,]\!\!\!]\,}
\def\dC{C\kern-6.5pt I}

\def\bw{\bar w}
\def\by{\bar y}

\def\bW{\overline W}

        \def\CC{{\cal C}}
        
\def\CG{{\cal G}}

        \def\CR{{\cal R}}

\def\un{{\bf 1}}

%xxxxxxxxxxxxxxxxxxxxxxxxxxxxxxxxxxxxx

%%%%%%%%%%%%%%%%%%%%%%% amsTEX characters %%%%%%%%%%%%%%%%%%

\def\IC{\relax\hbox{$\inbar\kern-.3em{\rm C}$}}

\input amssym.def
\input amssym.tex
\def\IZ{\Bbb Z}\def\IC{\Bbb C}
\def\gg{\goth g}

%%%%%%%%%%%%%%%%%%%%%%

%%%%%%%%%%%%%%%%%%%%%%%%%%%%%%%%%%%%%%%%%%%%%%%%%%%%%%%%%%%%%%%%%
%
%           Current Definitions
%

 \def\bgo{\overline{\goth g}} %%% (affine) Lie
        %%% Cartans

                   %%% (simple/real) roots
\def\Rr{\Delta^{\rm re}} \def\bP{\overline\Pi}
                      % horizontal Weyl vector
                      % horizontal roots 

                       %%% weight/root lattice
 
\def\FW{\Lambda}\def\fw{\overline\FW}       % (horizontal)
                                            %fund.weights 

\def\bW{\overline W} \def\tW{\tilde W}      %%% Weyl groups
\def\hw{w} \def\bw{\overline w}

                 % KW action on left
                       % horiz. KW group
                                 % group unit
\def\UNIT{\blacktriangleleft\kern-.6em\blacktriangleright}

                               %%% translations

\def\ts#1,#2{{\tt e}^{#1\zk#2}}

                %%% cyclic groups  

\def\iii{\iota}                           %%% the map `iota'

\def\CC{{\cal W}^{(+)}} \def\tC{\tilde\CC} %%% chambers

\def\UU{{\cal U}}                         %%% `fiber' over w_0

                          %%% supports
         %%% `hats' or `head tiles' 
   %#1#2{{\cal V}_{#1}^{(#2)}}   %%% supp of `Verma'
                 %%% supp of f.d.  module
\def\GG{{\cal G}}                 %%% supp of f.d. `module'
                          %%% ordinary Verma module
                %%% ordinary Verma module 
                                             %      of type #2

\def\cc{\chi} \def\bc{\overline\cc}       %%% characters

                                 %%% `horizontal' denominator
                           %%% `invariant' denominator
                        %%% generalized `denominator'
                                 % Kostant partition

\def\ml{m}                                 %%% multiplicities
\def\bm{\overline{m}}                 %   ordinary multiplicities
                               %%% fundamental characters

\def\CR{{\goth W}} 
                    %%% rings 

  %%% filtrations
  %%% filtrations
  %%% filtrations
  %%% filtrations

            %%% number rings, cones, etc.
\def\IZp{\IZ_{\ge 0}} 

\def\IQ{{\Bbb Q}}

  %%% triality

                          %%% level

        %%% dimensions

                    %\def\hk{{\hat\kmath}}
 
\def\bs{\backslash} 

\def\ol#1{\overline{#1}}

\def\PROP{\medskip\noindent{\bf Proposition}\ \ }

\def\endPROOF{\quad$\square$\medskip}

%%%%%%%%%%%%%%%%%%%%%%%%%%%%%%%%%%%%%%%%%%%%%%%%%%%%%%%%%%%%%%%%%

\hfuzz=10pt
\font\tfont=cmbx10 scaled\magstep2 
\font\ifont=cmmi10 scaled\magstep3
\font\iifont=cmti10 scaled\magstep1
\font\rrfont=cmr10 scaled\magstep2 
\font\ttfont=cmbx10 scaled\magstep3 
\font\sfont=cmbx10 scaled\magstep1 
\font\rfont=cmr10 scaled\magstep1 
\font\male=cmr9
\vsize=8.5truein \hsize=6truein
\nopagenumbers 
\topskip=0.5truein
\raggedbottom 
\abovedisplayskip=3mm 
\belowdisplayskip=3mm 
\abovedisplayshortskip=0mm 
\belowdisplayshortskip=2mm 
\normalbaselineskip=12pt 
\normalbaselines

%%%%%%%%%%%%%%%%%%%%%%%%%

\lref\AY{Awata H. and  Yamada Y., 
       Fusion rules for the fractional level $\widehat{sl}(2)$ 
algebra,         Mod. Phys. Lett. {\bf A7}, 1185--1195 (1992).}

\lref\FMa{Feigin B.L. and Malikov F.G., 
        Fusion algebra at a rational level and cohomology of 
           nilpotent subalgebras of $\widehat{sl}(2)$,
      Lett. Math. Phys. {\bf 31}, 315--325 (1994).} 

\lref\FGPa{Furlan P., Ganchev A.Ch. and Petkova V.B., 
             Quantum groups and fusion rule multiplicities,
              Nucl. Phys. {\bf B343}, 205--227 (1990).} 

\lref\FGP{Furlan P., Ganchev A.Ch. and Petkova V.B.,
        Fusion rules for admissible representations of affine 
        algebras:           the case of $A_2^{(1)}$, 
         Nucl. Phys. {\bf B518} [PM], 645--668 (1998)\semi
 Furlan P., Ganchev A.Ch. and Petkova V.B.,
         An extension of the character ring of $sl(3)$ and its quantisation.
		Comm. Math. Phys. {\bf 202 } 701--733 (1999). }

\lref\GPW{Ganchev A.Ch., Petkova V.B. and Watts G.M.T., A note
on decoupling conditions for generic level $\widehat{sl}(3)_k $
and fusion rules, 
%hep-th/9906139, 
Nucl. Phys. {\bf B571} [PM] 457--478 (2000).}
 
\lref\Hump{Humphreys J.M., {\it Reflection Groups and
           Coxeter Groups}, (Cambridge University Press, 1990).} 

\lref\K{Kac V.G., {\it Infinite-dimensional Lie Algebras}, third
        edition, (Cambridge University Press, 1990).}

\lref\KW{Kac V.G. and  Wakimoto M.,  
      Modular invariant representations of infinite-dimensional 
      Lie algebras and superalgebras,
    Proc. Natl. Sci. USA {\bf 85}, 4956--4960 (1988)   \semi 
 Kac V.G. and  Wakimoto M.,
         Classification of modular invariant 
                 representations of affine algebras,
               Adv. Ser. Math. Phys. vol {\bf 7}, pp. 138--177. 
(World Scientific, Singapore, 1989)         \semi
 Kac V.G. and  Wakimoto M., 
        Branching functions for winding subalgebras and tensor 
products, 
          Acta Applicandae Math. {\bf 21}, 3--39 (1990).}

\lref\Zh{Zhelobenko D.P., Compact Lie Groups and their Representations, 
(AMS, Providence, Rhode Island, 1973).}

\lref\MW{Walton M., 
       Fusion rules in Wess-Zumino-Witten models, 
           Nucl. Phys. {\bf B340}, 777--790 (1990). }

%%%%%%%%%%%%%%%%%%%%%%%%%%%%%%%%%%%%%%%%%%%%%%%%%%%%%%%%%%%%%%%%%%%%

%%%%%%%%% start of text %%%%%%%%%%%%%

\null 
\vskip 1truecm 
 
\centerline{{\ttfont Fusion Rings Related to}}
\vskip 2truemm
\centerline{{\ttfont  Affine Weyl Groups\footnote{${}^{\star}$}
{Talk given by V.B.P. at
``Lie Theory and Its Applications in Physics III'', Clausthal,  11-14 July, 
1999, to appear in the Proceedings,  eds. H.-D. Doebner et al.
 } }}
\vskip 1.5cm

\centerline{{\sfont P. Furlan$^{*,\dagger}$} 
~{\rfont and}~ {\sfont V.B. Petkova$^{**}$}} 

\vskip 0.5 cm 
 
\centerline{
 $^{*}$Dipartimento  di Fisica Teorica
 dell'Universit\`{a} di Trieste and }
\centerline{$^{\dagger}$Istituto Nazionale di Fisica Nucleare
(INFN), Sezione di Trieste} 
\centerline{Strada Costiera 11}
\centerline{34100 Trieste, Italy}

\centerline{$^{**}$Institute for Nuclear Research and
Nuclear Energy} 
\centerline{Tzarigradsko Chaussee 72}
\centerline{1784 Sofia, Bulgaria}
\vskip 1cm 

\centerline{\bf Abstract} 

\midinsert\narrower\narrower\male
The construction of the fusion ring of a quasi-rational CFT based
on $\widehat{sl}(3)_k\,$ at generic level $k\not  \in \IQ\,$ is
reviewed.  It is a commutative ring generated by formal
characters, elements in the group ring $\IZ[\tW]$ of the extended
affine Weyl group $\tW$ of $\widehat{sl}(3)_k\,$.  Some partial
results towards the  $\widehat{sl}(4)_k\,$ generalisation of this
character ring are presented.
\endinsert

\vskip 1cm 

\noindent
{\tfont 1 \ Introduction}
\vskip 0.3cm 

\noindent 
Describing the fusion rules of the conformal field theory (CFT)
based on the  fractional level
 admissible representations of the affine KM algebras \KW\
remains an open problem in general. It has been solved
in the simplest case of $   \widehat{sl}(2)_k\,$ \AY\
by a direct solution of the singular vector decoupling equations,
the result being confirmed by a more abstract analysis in \FMa.
In \FGP\ a general approach was initiated and illustrated on the
first example 
with nontrivial fusion multiplicities, the case $   \widehat{sl}(3)_k\,.$  
The  problem is treated  
starting first with
 a   quasi-rational CFT  described by an infinite
class of ``pre-admissible'' representations of the affine
algebra with generic $k\not  \in \IQ\,$ and
highest weights labelled by a certain subset
of the
extended affine Weyl group. 
The fusion rules of this theory
provide the ``classical'' counterpart
of the fusion rules of the fractional level admissible  CFT.
This is analogous to the relation  between the  CFT
based on the integrable representations and its
``classical'' counterpart, 
 the ``pre-integrable''
CFT  described by an infinite set of representations
with  $k\not  \in \IQ\,$
and  highest weights given
by integer dominant weights of the horizontal subalgebra.
In that case
the fusion rule multiplicities 
of the generic level 
%quasi-rational 
CFT are given simply by the
 tensor product multiplicities of the finite dimensional 
representations of the horizontal algebra. The standard
formal characters of these
representations, 
%formal exponentials -- 
elements of the group ring of the
group of translations, serve as characters of the fusion algebra
of the 
 affine algebra representations .
These ``classical'' data are partially incorporated in the integrable
representations 
CFT, namely the   
 formula for
the  fusion rules derived in
\K,\MW,\FGPa\
 is based on 
the notion of   weight diagram
 (a support of a finite dimensional module),
and
furthermore the  characters of the integrable fusion algebra
 can be recovered from the standard 
characters by certain quantisation procedure.  
Similarly all these structures need to be generalised in order
to describe the fractional level admissible CFT.
In particular it turns out that the standard formal characters
are  replaced by some elements in the group ring of the
(extended) affine Weyl group. 

In the next section
 we  review the 
main steps in \FGP\ and in the last section we turn to the case of 
$\widehat{sl}(4)_k$.

\vskip 1cm 

\noindent
{\tfont 2 \ General setting and the case of} 
~$\widehat{\hbox{\ifont sl}}{\hbox{\rrfont (3)}}_{\hbox {\iifont k}}$

\vskip 0.3cm
 
\noindent
Let $k\not  \in \IQ\,$ and consider the
subset $\tC$  of the
extended Weyl group $\tW$, defined as
\eqn\Ida{
\tC:=\{y\in \tW\,|\, y(\za_i)\in \Rr_+\ \ {\rm for}\ \forall
\za_i\in \bP\}\,,
}
where  $\tW=\bW\ltimes t_P=
W\rtimes A\,,$  $t_P$ is
the subgroup of translations in the weight lattice $P$ of $\bgo=sl(n)\,,$
$\bW$ and $W$ are the Weyl groups of $\bgo$ and $\gg=\widehat{sl}(n)_k$
 respectively,
and $A$ is the cyclic subgroup of $\tW$ which keeps invariant the set of
simple roots $\Pi=\{\za_0\,,\za_1,\dots, \za_{n-1}\}\,$ of $\gg$. In \Ida\
 $\Rr_+$  is the set of real positive roots of
$\gg$ and  $\bP$ is the  set of simple roots of $\bgo$.

Denote $\CC=\tC\cap W$, then $\tC=\cup_{a\in A}\, a\CC$.
The subset $\tC$($\CC$) is a fundamental domain  (a ``dominant chamber'')
in $\tW$($W$) with respect to the right action of $\bW$ \FGP\
and the subset $\tC\cdot k\zL_0 $   of  weights 
 (or, equivalently, the subset $\tC\subset \tW$ itself)
 labels the highest weights $\Lambda$ of
 maximally reducible  Verma modules of $\gg$. Indeed for 
$\Lambda=y\cdot k\Lambda_0$ and $\beta =y(\za)\,,$ s.t. $y\in \tC$,
the Kac-Kazhdan singular vector criterion holds true for any positive
root $\za$ of $\bgo.$
The   Kac-Kazhdan reflections can be identified with the right action
of $\bW$ on $\tW\,,$ i.e., 
$$w_{y(\za)}\cdot \Lambda =
 y w_{\za}\cdot k \Lambda_0\,.
$$ 
Here $\zL_0$ is the
fundamental weight of $\gg$ dual to the affine root $\za_0\,,$
 and the  shifted action of $\tW$  is given by 
$\hw\cdot\zL=\hw(\zL+\zr)-\zr\,,$ $\rho $ being the Weyl vector of $\gg$. 

 Introducing a map $ \iii$ of $\tW$  into the root
lattice $Q$  of $\bgo$  
\eqn\mi{
 \iii: \tW\ni y=\by t_{-\zl} \
                                \mapsto \
     n\,\zl + \by^{-1}\cdot 0
	 \in Q\,,
}
with the properties
\eqn\tlg{\eqalign{
\iii(x y ) & = \ol{y}^{-1}(\iii(x)) + \iii(y) \,,\cr
 \iii(y\bw) & =
 \bw^{-1} \cdot\iii(y)\,, 
\quad \bw  \in\bW\,,
}}
one can express $\tC$ alternatively as
\eqn\If{
\tC=\{y\in \tW\, |\, \iii(y) \in P_+\}\,,
}
where $P_+= \oplus_{i}\ \IZp\ \fw_i\,,$ $\fw_i$ being the
fundamental weights of $\bgo$. The $\bgo$ Verma modules of highest weight
$\iota(y)$ are reducible iff the corresponding $\gg$ Verma modules
 of highest weight $\Lambda=y\cdot k\Lambda_0$ are reducible.

With any $y\in \tC$ 
an element of the group
ring  $\IZ[\tW]\,$ of $\tW$,
 a formal ``character'', is associated
\eqn\res{
\cc_{y}\,
  =  \sum_{z\in\tW\,,\, zy^{-1}\in W} \ml_{z}^{y} \ z\,,
}
and extended to $\tW$,
\eqn\resa{
	\cc_{y\bw}:
 =\det(\bw)\,\cc_{y}\,, \quad y\in \tC\,, \ \bw\in
\bW\,,
}
where the integer coefficients $ \ml_{z}^{y}$ are defined in the $n=3$ case
as
\eqn\mt{
\ml_{z}^{y}
=\bm^{\iii(y)}_{\iii(z)}\,,
}
  $\bm^{\iii(y)}_{\iii(z)}$ being the standard multiplicity of the
weight $\mu=\iii(z)$ of the finite dimensional representation of
$sl(3)$ of highest weight $\lambda=\iii(y)$.
One introduces the  notion of a generalised ``weight diagram'', 
$\GG_{y}=\{z\in \tW \,|\, \ml_{z}^{y}\ne 0\}\,,$ interpreted as
the support  (i.e., a set of weights with their multiplicities)
 of a ``finite dimensional module'' of highest weight $y$.
These generalised  weight diagrams are explicitly
determined by \mt\ and thus have the structure of the  weight
diagrams $\zG_{\iii(y)}$ of 
 triality
zero $sl(3)$ representations,
 with the  weights 
$\mu\not\in\ $Im$(\iii)$ excluded. In particular we refer to $y$
as a highest weight element in the diagram $\GG_{y}$ or in the
character $\cc_y$.

The characters \res\ give rise to a commutative ring with identity,
interpreted as an extension of the 
ring
of characters of the finite dimensional representations of $sl(3)$.
The ring contains a subring generated by elements $\cc_y$
 labelled by $y\in \CC=\tC\cap W$, each represented as a polynomial of
three ``fundamental''  characters, 
 $\cc_{w_0}\,, \cc_{w_{10}}\,,\cc_{w_{20}}\,,$ subject
to one relation. To the elements of $A$ correspond  "simple
currents", $\cc_a =a\,, a\in A\,,$
\eqn\sc{
\cc_a\ \cc_y
=\cc_{ay}\,,
}
so that 
the fusions \sc\ with  the
generator $\gamma=\cc_{\gamma}$ of $A$  recover 
all $\cc_y\,,$ $y\in \tC$. 

Having a notion of a generalised weight diagram,
a  formula for the fusion rule multiplicities,
 generalising the Weyl-Steinberg (W-S) formula, can be derived,
\eqn\ws{
     \cc_x\ \cc_y 
 = \sum_{z\in {\cal G}_x} \ml^{x}_{z} \ \cc_{zy} 
  = \sum_{z\in \tC} N^{z}_{x,y} \ \cc_{z} \,,
}
\eqn\wsa{
N^{z}_{x,y}
 = \sum_{\bw\in\bW} \, \det(\bw)\ \ml^{x}_{z\bw y^{-1}} \,.  
}
The second equality in \ws\ is derived as for the usual
$sl(n)$ characters, using the symmetry in \resa\ and the fact that
$\CC$ ($\tC$)
is a fundamental domain in $W (\tW$); the summation in $z$ in
the last term 
runs effectively over the 
shifted weight diagram $\CG_x\,y \cap\tC\,$ of `shifted highest
weight' $xy\,.$

Using the properties \tlg\ of the map $\iota$ the fusion coefficients
 \wsa\ can be also expressed in terms of
the structure constants $\bar{N}^{\iii(z)}_{\iii(x)\,\iii(y)}$ 
 of the $sl(3)$ character ring
\eqn\strci{
N^{z}_{x,y} = \bar{N}^{\iii(z)}_{\iii(x)\,\iii(y)} \,.
}

The generic fusions for the four generators 
$\cc_{w_0}\,, \cc_{w_{10}}\,,\cc_{w_{20}}\,, \cc_{\zg}$ were confirmed
in \FGP, \GPW\
to reproduce the CFT fusion rules.
Finally a ``quantisation'' of  these formal characters leads  to the
finite set of characters of the fusion algebra of rational level $k=3/p$
admissible representations, see \FGP\ for more details. 
\bigskip

It turns out that the definition of the generalised weight diagram
 in \FGP,
based on \mt,  has to be modified
in the $sl(4)$ and presumably in  all higher rank cases, since it is not
consistent with the Weyl-Steinberg formula \ws, as can be checked in
simple examples, see next section.
This in particular implies that the 
formula
\strci, alternative to \wsa\ in the $n=2,3$ cases,  does not hold
true in general.
Typically the dimension of the ``module'' prescribed by \mt\ is higher
than what one obtains by fusion, applying the W-S type formula
\ws\ and adopting 
the definitions \res,\mt\ for a subset of  smaller dimension
modules,   see below.

\vfil\eject
\vskip 1cm

\noindent
{\tfont 3 \  The case of}  
~$\widehat{\hbox{\ifont sl}}{\hbox{\rrfont (4)}}_{\hbox {\iifont k}}$
\vskip 0.3cm
\noindent
We start with recalling some standard notation for the algebra 
$\gg=\widehat{sl}(4)_k$ and its Weyl group. 

The Weyl group of $\bgo=sl(4)$ consists of $24$ elements and its 
Cayley graph is drawn in Fig. 1. It is an octahedron with truncated
tips, so it has
$8$ elementary hexagons and $6$ 
squares. They correspond to the Artin relations 
$w_iw_{i+1}w_i=w_{i+1} w_i w_{i+1}\,,$ $i=1,2$ and $w_1 w_3=w_3 w_1$ 
 respectively. 
A finite part of the 
Cayley graph of the affine Weyl group $W$ is depicted  in Fig.2. with the
unit element $\un$ denoted by $\star$. It is made
of elementary truncated octahedrons isomorphic to the graph
 representing $\bW$ with the
generators $w_i\,, i= 1,2,3 $ replaced by $a w_ia^{-1}\,, a\in A\,$ 
(i.e., by
$(w_2\,,w_3\,,w_0)\,,$ etc).
Note that unlike the $sl(3)$ Cayley graphs exploited in \FGP\
here the graphs are not drawn as  geometric graphs with angles and lengths 
consistent with the projection to the root diagram of $sl(4)$
 defined by the map
$\iota$ in \mi. Rather each elementary truncated octahedron isomorphic to
 the finite Cayley 
graph
of $\bW$ is ``squashed'' so that two adjacent hexagons originally meeting at an
angle $2$ arctg $\sqrt 2$ now lie on one and the same plane, cf. Fig. 1. Thus  
the whole $W$ graph is
organized into layers (part of a layer is clearly seen in Fig. 2),
 and if we single out a positive direction along anyone of
the edges belonging to the basis of the two opposite pyramids forming an
octahedron, we
see that within any layer we can distinguish three kinds of vertices:
 those ones
only connected to vertices of the preceding layer ( denoted by a white circle
), those ones only connected to vertices of the following layer (denoted by a
black circle) and those ones only connected to vertices of the same layer 
(denoted by no circle).

\fig{}{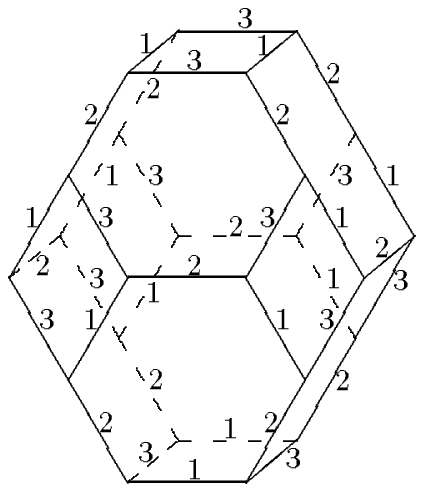}{6cm}

 The cyclic subgroup $A$ of $\tW$ is generated by 
 $\zg= t_{\fw_1}\, \bar{\zg}$, where $ \bar{\zg}=w_{123}=w_1\,w_2\,w_3$
  is a Coxeter element in $\bW$ generating  the 
cyclic subgroup $\bar{A}$ of $\bW$.
  One has
$\zg(\za_j)=\za_{j+1}=\zg^{j+1}(\za_0)$ for $j=0,1,2,3,$
identifying $\za_4\equiv\za_0$, and accordingly $A$ gives an automorphism
of $W$, 
$w_{\za}\rightarrow \zg \, w_{\za} 
\zg^{-1}=w_{\zg(\za)}\,,$ $\za\in \Pi$.  The 
 Cayley graph of $\tW$ can be thought of  as a 4-sheeted covering
of the graph of $W$ with, for example, the ``fiber'' over the
edge `0' connecting the vertices $\un$ and $\hw_0$ being the set
$\{A, A\,\hw_0\}$.  In Fig. 2 we have drawn schematically part of  the
first sheet $\CC$ of the fundamental domain $\tC$ with
the first layer and  
part of the interior structure indicated;
e.g., the second layer starts with the point $w_{210}$ as an ``origin''
and part of its edges are indicated with dashed lines.
 We have also indicated
the position of the $\iota$ - images of some of the elements in $\CC$,
but we have to caution the reader that among the black and white points,
visually appearing on the second fundamental
axis $\fw_2\,,$ only the white circled
elements have in reality their $\iota$-images lying on this
axis with coordinates $(0,4 n,0)$, $n=0,1,\dots$.

\fig{}{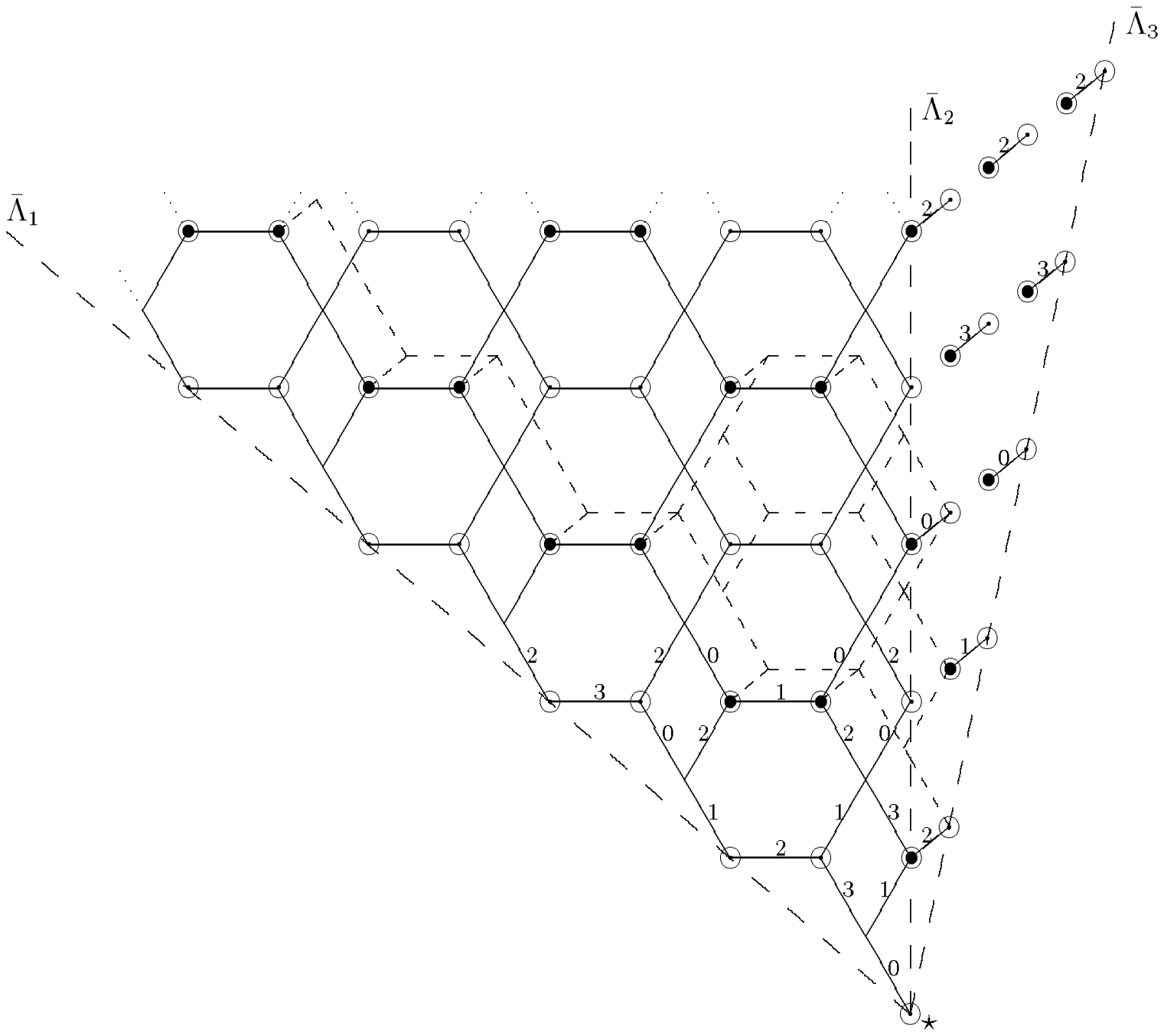}{12cm}

Define the subset of $\tW$
\eqn\us{
\UU=\{A\,, A\,w_0\,, A\,w_{10}\,, A\,w_{30}\,, A\,w_{310}\,,
A\,w_{2310} \}\,.
}
Its projection $\bar{U}$ onto the subgroup $\bW$ gives the right
cosets of  $\bar{A}$.
It is easy to show that the chamber $\tC$ is represented as
$\tC=\UU\,t_{-P_+}$. 

We define $4$ term ``class'' elements in the group ring $\IZ[W]$  
\eqn\df{
F_{rst...}\equiv F_{w_{rst...}}:= \sum_{a\in A}aw_{rst...}a^{-1}\,,
\quad w_{rst...}\in W\,.
}
We shall furthermore  denote $"F"=F/2$ 
in the cases when $F$ contains two terms
with multiplicity 2, an example is provided by $"F_{13}"= w_{13} +w_{02}
="F_{20}"$. Note that
since $\iota(a)=0$ 
for any $a\in A$, each of the terms in a given $F_y$ is mapped to
one and the same  $\bW$ orbit in the $sl(4)$ weight diagram
and  their images form a $\bar{A}$ orbit. 
Apparently $F_y=F_{aya^{-1}}$ for
$\forall
 a\in A$ while for the product of two such elements we have, with 
 the multiplication  inherited
from the multiplication in 
$\IZ[W]$, 
\eqn\prf{
F_x\ F_y = \sum_{a\in A}\, F_{xaya^{-1}}=\sum_{a\in A}\, F_{axa^{-1}y}\,.
}
In general $F_x\ F_y \not = F_y\ F_x$, but e.g., $F_0\,,
F_{30}+"F_{13}"\,, F_{10}+"F_{13}"\,, $
commute between themselves.

We shall now introduce a finite set of formal characters
$\cc_y\,, y\in \CC\,,$ as in $\res$,
for all of which we will adopt the definition  $\mt$.
In employing the map \mi\ and comparing with the standard $sl(4)$ weight
diagrams one can use the 
recursive  formula 
 for the multiplicity of a weight 
 $\mu$ (see, e.g. \Zh{})
\eqn\fre{
m_\mu =\; -\sum_{\overline{w}\in\overline{W};\,\overline{w}\not = {\bf 1}}
{\rm det}(\overline{w})\  m_{\mu +\rho - \overline{w}(\rho ) }\,,
}
with the weights in the r.h.s.  strictly greater than $\mu$.

We have  for $y\in \CC$ and  of 
length $l(y) \le 3$
\eqna\Ia
$$\eqalignno{
\cc_{w_0}& = 3 + w_0 + w_1 + w_2 + w_3 \quad\equiv 3 + F_0\,, \qquad
\iota(w_0)=(1,0,1)\,,\cr
\cc_{w_{10}} & = 3 + 2F_0 + "F_{13}"  + F_{10} \,,\ \ \ \qquad
\iota(w_{10})=(0,1,2)\,,\cr
\cc_{w_{30}}& = 3 + 2F_0 + "F_{13}"  + F_{30} \,, \ \ \  \qquad
\iota(w_{30})=(2,1,0)\,,\cr
\cc_{w_{230}} & =\cc_{_{\zg\, t_{-\fw_1}}}= 1 + F_0 + "F_{13}" +
 F_{30} + F_{230}\cr
&= 1 + F_0 + "F_{13}" + F_{30} + \zg\bc_{\fw_1}\,, \qquad
\iota(w_{230})=(4,0,0)\,, &\Ia{}\cr
\cc_{w_{210}}& =\cc_{_{\zg^3\, t_{-\fw_3}}}= 1 + F_0 + "F_{13}" + F_{10} +
 F_{210}\cr
&= 1 + F_0 + "F_{13}" + F_{10} + \zg^3\bc_{\fw_3}\,, \qquad
\iota(w_{210})=(0,0,4)\,,\cr
\cc_{w_{130}} & = 7 + 5F_0 + 4"F_{13}" + 2(F_{30} + F_{10}) +
F_{121} + (F_{130} + F_{213}) \,,\cr
}$$
(where $\iota(w_{130})=(1,2,1)$) of dimension $7,17,17,15,15,63$
 respectively. Here $\bc_{\fw_i}$ are 
standard formal $sl(4)$ characters, i.e., elements in the group ring
$\IZ[t_{P}]$. The
 terms  combined in parentheses have $\iota$ -- images belonging to
one and the same orbit of $\bW$. To each of these characters we associate
 a weight diagram, which can be   identified with a finite subset
of the Cayley graph of $W$, the identity term being identified with
the identity element in $W$, with multiplicities assigned to each
vertex. E.g., the diagram ${\cal G}_{w_0}$ for 
$\cc_{w_0}$
consists of the identity vertex in the center, with assigned multiplicity $3$
 and the four vertices connected to it by the generators 
$w_i\,.$

Exploiting \prf\ one obtains by a
direct computation
  the  relations
\eqn\fr{\eqalign{
\cc_{w_0}\ \cc_{w_0} &= 1 + 2\, \cc_{w_0} + \cc_{w_{10}} + 
     \cc_{w_{30}}\cr 
\cc_{w_0}\ \cc_{w_{10}} &= \cc_{w_0} + 2\,\cc_{w_{10}}  + 
     \cc_{w_{130}}    + \cc_{w_{210}}\cr
\cc_{w_0}\ \cc_{w_{30}} &= \cc_{w_0} + 2\,\cc_{w_{30}} + \cc_{w_{130}} + 
     \cc_{w_{230}}\,.
}}
These relations are consistent with the W-S formula \ws,\wsa. E.g., 
for the first fusion  we get following \ws\
$$\cc_{w_0}\ \cc_{w_0}=
\sum_{z\in {\cal G}_{w_0}}\, m_z^{w_0}\, \cc_{z w_0}=
 1 + 3\, \cc_{w_0} + \cc_{w_{10}} + 
     \cc_{w_{30}} +\cc_{w_{20}}
\,.
$$ 
The last term
$\cc_{w_{20}}=\cc_{w_{02}}$ has a 
highest weight $w_{20}$ which does not belong
to the dominant chamber $\tC$ and according to  \resa\

$\cc_{w_{20}}=\cc_{w_0 w_2}=$ det$(w_2)\, \cc_{w_0}=- \cc_{w_0}$, thus 
$\cc_{w_{20}}+3\,\cc_{w_0}=
2\,\cc_{w_0}$, 

\noindent
recovering the first equality in \fr\ with fusion multiplicity
$N_{w_0 w_0}^{w_0}=2$.

The relations \fr\ imply that given the characters labelled
by  $w_0\,, w_{10}\,, w_{210}\,$ (or by
$w_0\,, w_{30}\,, w_{230}\,$), one can generate the remaining 
characters with highest weight $y$ of length $\le 3$. 
In the next step there are $4$
characters $\cc_y$ with $y\in\CC$ of length 4, and all of them
are recovered recursively fusing according to \ws\
and comparing with the result of the explicit multiplication, e.g., 
\eqna\fl
$$\eqalignno{
\cc_{w_0}\ \cc_{w_{230}} &=\cc_{w_{30}}+\cc_{w_{230}}+\cc_{w_{1230}}\,,\
\ \ \qquad \iota(w_{1230})=(5,0,1)\,,\cr
\cc_{w_0}\ \cc_{w_{210}} &=\cc_{w_{10}}+\cc_{w_{210}}+\cc_{w_{3210}}\,,\
\ \ \qquad \iota(w_{3210})=(1,0,5)\,, &\fl{}\cr
\cc_{w_{10}}\ \cc_{w_{30}} &= 1+2\, \cc_{w_{0}}+\cc_{w_{10}}+ \cc_{w_{30}}+
\cc_{w_{130}}+\cc_{w_{2130}}\,,
\quad \iota(w_{2130})=(2,2,2)\,,\cr
\cc_{w_{10}}\ \cc_{w_{10}} &= \cc_{w_{10}}+ \cc_{w_{30}}+ \cc_{w_{210}}+
2\, \cc_{w_{130}}+ \cc_{w_{3210}}+ \cc_{w_{0130}}
\,,\cr
\cc_{w_{30}}\ \cc_{w_{30}} &= \cc_{w_{10}}+ \cc_{w_{30}}+ \cc_{w_{230}}+
2\, \cc_{w_{130}}+ \cc_{w_{1230}}+ \cc_{w_{0130}}\,.
}$$
The last character, $\cc_{w_{0130}}=\cc_{_{\zg^2\, t_{-\fw_2}}}\,,$ 
with $\, \iota(w_{0130})=(0,4,0)\,,$
contains a term $F_{0130}+ "F_{0213}"=\zg^2\, \bc_{\fw_2}$.

While the characters  $\cc_{w_{1230}}\,,\cc_{w_{3210}}\,,
\cc_{w_{0130}}\,,$  obtained by the 
fusions in \fl{}\ coincide with the expressions prescribed by
 \mt, the character $\cc_{w_{2130}}\,,$
with $\, \iota(w_{2130})=(2,2,2)$,
provides the first example in which 
formula \mt\
fails. Indeed the  expression obtained by the fusion
is ``smaller', with some terms missing, or multiplicities
decreased, effectively implying that the corresponding
weight diagram is a subset of the one determined by
\mt.  Using the above fusion
the character (corresponding to the ``true'' weight diagram) is given by
$$\eqalign{
\chi_{w_{2130}} &= 11 + 9F_0 + 8\, "F_{13}" + 5 F_{12} + 5 F_{21} + 3F_{121} +
3(F_{213} + F_{132})\cr 
&+ 2F_{123} + 2F_{321} +  (F_{1213} +  F_{1232}) +
(F_{1321} +F_{2321}) + "F_{0213}"  + F_{2130}\,. 
}$$

In the Table below we compare the two weight diagrams, the one prescribed
by
\mt, and the ``true'' one, obtained by fusion. In the first column the
dominant weights on each 
Weyl orbits of the $sl(4)$ representation of highest weight
$(2,2,2)=\iota(w_{2130})$ are listed,
with 
the multiplicities recalled in the second 
column.
We select those $\bar{A}$ orbits on each $\bW$ orbit
which lie in the image  of the map $\iota$ ; e.g., there are two such
$\IZ_4$ orbits with 8 elements  in the first $\bW$ orbit generated by the
highest weight $\bL =(2,2,2)$, etc. 
The 
third column contains the corresponding ``would be'' terms in the
character, while the last column contains the correct multiplicities,
 which correspond to the explicit expression  for
$\cc_{w_{2130}}$ above obtained by fusion.

\medskip

\halign{
$#$&\qquad $#$&\qquad $#$&\qquad $#$&\qquad $#$&\qquad $#$&\qquad $#$\cr
dominant\, weight & {}
& sl(4)
&wrong\, contr. &\,\,\, true
\cr
{} & {} 
&mult. 
&to\, \chi_{w_{2130}} &\  mult. 
\cr
{}&{}&{}&{}&{}&{}
\cr
\lla  & =\, (2,2,2) 
& 1 & F_{2130}+F_{121321}&\,\,\, 1|0
\cr
\lla -\za_1 & =\, (0,3,2) 
& 1 & F_{21321}&\,\,\, 0
\cr
\lla -\za_2  & =\, (3,0,3) 
& 1 & F_{12321}&\,\,\, 0
\cr
\lla  -\za_3 & =\, (2,3,0) 
& 1 & F_{12132}&\,\,\, 0
\cr
\lla -\za_1 -\za_3 & =\, (0,4,0) 
& 1 &F_{2132}+"F_{0213}"&\,\,\, 0|1
\cr
\lla -\za_1-\za_2  & =\, (1,1,3) 
& 2 & F_{1321}+F_{2321}&\,\,\, 1|1
\cr
\lla -\za_2 -\za_3 & =\, (3,1,1) 
& 2 & F_{1213}+F_{1232}&\,\,\, 1|1
\cr
\lla -2\za_1-2\za_2  & =\, (0,0,4) 
& 3 & F_{321}&\,\,\, 2
\cr
\lla -2\za_2-2\za_3  & =\, (4,0,0) 
& 3 & F_{123}&\,\,\, 2
 \cr
\lla -\za_1-\za_2 -\za_3 & =\, (1,2,1) 
& 4 & F_{132}+F_{213}&\,\,\, 3|3
\cr
\lla -\za_1-2\za_2 -\za_3 & =\, (2,0,2) 
& 5 & F_{121}&\,\,\, 3
\cr
\lla -2\za_1-2\za_2 -\za_3 & =\, (0,1,2) 
& 7 &F_{21}&\,\,\, 5
\cr
\lla -\za_1-2\za_2 -2\za_3 & =\, (2,1,0) 
& 7 &F_{12}&\,\,\, 5
\cr
\lla -2\za_1-2\za_2 -2\za_3 & =\, (0,2,0) 
& 10 &"F_{13}"&\,\,\, 8
\cr
\lla -2\za_1-3\za_2 -2\za_3 & =\, (1,0,1) 
& 12 &F_0&\,\,\, 9
\cr
\lla -3\za_1-4\za_2 -3\za_3 & =\, (0,0,0) 
& 15 &1&\,\,\, 11
\cr
}
\bigskip

{}From the Table we see  one of the reasons why \mt\ comes
into contradiction with the fusion based on \ws: there are 
elements of higher length than the length of
the  highest weight element $y=w_{2310}$ which have images in the 
 weight diagram of the $sl(4)$ representation $\iota(y)$.
Indeed, according to \mt,
the  longest element $w_{121321}$ of $\bW$
 contributes to the character along with the
highest weight element $w_{2130}$,
since both $\iota(w_{121321})$ and $(2,2,2)=\iota(w_{2130})$
appear in the outer Weyl orbit in the $sl(4)$  weight diagram.
 On the other hand 
$F_{121321}$,  looked
as a constituent of $w_{121321}$,
cannot be produced multiplying 
two elements of
$\bW$ with a sum of lengths giving $4$.

 \bigskip

\noindent

Let us denote by ${\cal F}$ the set of the
first five characters in
\Ia{}\ with the identity $\cc_{\un}=\un$ added,
\eqn\bse{
{\cal F}=\{\un\,, \cc_{w_0}\,,\cc_{w_{10}}\,,\cc_{w_{30}}\,,
\cc_{w_{210}}\,,\cc_{w_{230}}\}\,.
}
The elements in \bse\ commute between themselves and 
 are restricted by two algebraic relations
implied by \fr.
Then we have

\PROP{\it For any $y\in \CC$ there is a formal character $\cc_y$ of 
the form of \res,
which is 
obtained recursively, 
using \ws, 
as a polynomial
of the elements in \bse.}
\medskip
%%%%%%%%%%%%%
\blank{
The resulting
polynomial ring generated by   $\cc_y\,,$ $y\in \CC$,
to be denoted $\CR$,
is a commutative subring  of $\IZ[\tW]$.
}
%%%%%%%%%%%%%%
Let us sketch the proof of the proposition. 

%%%%%%%%%
\blank{
First assuming that a character $\cc_y$ is represented as a polynomial
in the elements of ${\cal F}$ one  can prove
that the W-S formula \ws\ holds with $\cc_x$ replaced with any 
of the ``fundamental'' characters in \bs. The proof is analogous to the
second proof of Lemma 4.5 in \FGP\ and will be skipped here.
}

We have to prove that 
 by a proper fusion, identifying  ${\cal G}_x$ with the
weight diagram of some of the elements in ${\cal F}$, 
the r.h.s. of \ws\ always produces one and
 only one ``new'' character with highest weight  in $\tW$. 
It is useful to draw the weight diagrams of the basic elements in ${\cal F}$.
The shifted diagram ${\GG_x} y$ is visualised
 identifying the center of the
weight diagram  ${\GG_x}$, i.e., the identity element in any of
the first five characters in \Ia{},
with the point $y\in\tC$, which locates
${\GG_x}$
generically 
in the chamber $\tC$, each vertex counted with the corresponding multiplicity.
 Whenever some points of the shifted
weight diagram appear outside of $\tC$ their contribution is 
cancelled,
using the right action of $\bW$ in \resa, i.e., the weight multiplicities are 
replaced by the fusion multiplicities \wsa\ as in the last equality in \res.

We first consider the first layer in Fig. 2. Its vertices form the subset
$\UU\, t_{-n_1 \fw_1-n_2 \fw_2}\cap \ W$ of $\CC$.
\medskip
We can split this layer into a disjoint union of even  and odd
 horizontal ``floors'', to be denoted $T_{2j}$ and $T_{2j+1}$, $j=0,1,2\dots$,
 containing $2j+1$ and $j+1$ elements respectively.
 E.g., $T_0 =\{\un\}\,,$ $T_1=\{w_0\}\,,$ $ T_2=\{w_{230}\,, w_{30}\,,
 w_{10}\}\,,$ $T_3=\{w_{1230}\,, w_{130}\}\,,$
 $T_4=\{w_{301230}\,, w_{01230}\,,w_{21230}\,, w_{2130}\,, w_{0130}\}\,,$ etc.
Their $\iota$ - images
 are given by $\iota(T_{2j})=\{4j\,\fw_1\,,4j\fw_1-\za_1\,, 
4j\,\fw_1-2\za_1+\za_3\,,4j\,\fw_1-3\za_1+\za_3\,,4j\,\fw_1-4\za_1\,,
4j\,\fw_1-5\za_1+\za_3\,,\dots, 4j\,\fw_1-2j\,\za_1$, for $j$-even, (or,
$4j\,\fw_1-2j\,\za_1+\za_3$ for $j$ -odd)  $\}\,,$
 $\iota(T_{2j+1})=\{4j\,\fw_1+\theta\,,4j\,\fw_1+\theta-2\za_1\,,
4j\,\fw_1+\theta-4\za_1\,, \dots, 4j\,\fw_1+\theta-2j\,\za_1\}\,.$
 
The first three floors contain
 only elements labelling the highest weights of the fundamental
 set ${\cal F}$. 
 We shall identify them with the corresponding characters.
 Assume that for a given
 $j\ge 1$ all characters 
 in $T_l$ with $l\le 2j$
  are generated. Then starting from the ``white'' circles in
 $T_{2j}$ we recover fusing  by $w_0$ all
 the elements of the next floor $T_{2j+1}$,
  by $w_{10}$ -- all ``black'' circles in $T_{2j+2}$, and by $w_{30}$ --
  all  white circles in $T_{2j+2}$ with the exception of the utmost left one,
   with an $\iota$ image $(4j+4)\fw_1$; 
the latter is recovered by $\cc_{w_{230}}$ starting from
  the utmost left white circle in $T_{2j}$ mapped to
$4j\,\fw_1$. All of these fusions contain only
  one new element,  the remaining terms corresponding to elements in $T_l$
  with smaller $l$. 
  In this way one generates all the characters corresponding to the
  points on the first layer. 

 %%%%%%%%%%%%%
\blank{ 
In the next step we show that once we know the
  characters for each layer ${\cal L}_i$ with $i\le k$ for a given $k$, 
  we recover the layer ${\cal L}_{k+1}$. Indeed, starting again
from a black 
  circle on ${\cal L}_{k}$ and using the basic weight diagrams in \Ia\
  we get that all white circles on ${\cal L}_{k+1}$ are recovered by
  $w_0$, then all simple vertices are recovered by
  $w_{10}$, 
and finally all black  vertices  -- 
  by fusing with $w_{210}$.
 }
%%%% 

It remains to repeat the steps 
for any consecutive layer since any layer repeats the 
structure of the first one with the only
 difference that there are edges going backwards, i.e.,
producing characters
 already generated in the previous steps. What we need is a starting
point replacing the identity element, e.g., for the second layer
this point is $w_{210}$, the corresponding character being in the
fundamental 
set \bse. Then the first three floors are given by
$T^{(2)}_0=\{w_{210}\}\,, 
$ $T^{(2)}_1=\{w_3 w_{210}\}\,,$ generated by $\cc_{w_0}$ (recall that
$F_0=F_3$),
and $T^{(2)}_2=\{w_{123} w_{210}\,,
w_{23} w_{210}\,,w_{03} w_{210}\}\,, $ generated by $\cc_{w_{230}}\,,
 \cc_{w_{30}}\,,
\cc_{w_{10}}$ resp.,  acting on $\cc_{w_{210}}$; recall that
$F_{230}=F_{123}$, etc.). 
The starting points of the next layers
are depicted as white circles with images on the  $\bL_3$ axis.
Once the preceeding layer is recovered, each of the characters
labelled by these points
is produced starting from
the preceeding one and applying $\cc_{w_{210}}$. 
\endPROOF

The Proposition implies that comparing the result of the direct
 multiplication
 in the l.h.s of \ws\ with the r.h.s., which contains
at every recursive step only one ``new'' character, the latter
is obtained explicitly, i.e., the multiplicities in \res\ are determined.
It is also clear from the construction
that the element in $\IZ[W]$ corresponding to
the  highest weight $y$ of  $\cc_y$, as well as the full term
$F_y$,  appears with multiplicity one.
However the nonnegativity of the general multiplicities in \res\
 remains to be proved. We have checked  this for
up to length 6  highest weight words. All the examples confirm
the result for the character  $\cc_{w_{2130}}$ described above
and suggest that 
the relevant definition of the weight diagrams generalising
the one in the  $\widehat{sl}(3)_k$ case has to be based again on 
the correspondence via the map $\iota$
to the standard $sl(4)$ weight diagrams. Namely a term $F_z$ is present in
$\cc_y$
 only if $\iota(z)\in \Gamma_{\iota(y)}$, however in general 
$m_z^y\le \overline{m}_{\iota(z)}^{\iota(y)}$. 
We recall
that in the $\widehat{sl}(3)_k$ case the  multiplicities $m_z^y$
 were derived starting from  supports of
 generalised ``Verma modules'' and prescribing their multiplicities
by comparing,  via the map $\iota$, with the  multiplicities
of the $sl(3)$ Verma modules determined by the Kostant partition function;
 this led to a generalised Weyl formula for the  characters.
Thus the problem  can be reformulated
 as the problem
of finding proper definition for the generalised Verma module supports.

We conclude with the remark that there is a sufficient evidence that
as in the $\widehat{sl}(3)$ case, the $\widehat{sl}(4)_k$ fusion
 ring is an extension  of the  $sl(4)$
character ring $\overline{\CR}$. 
Indeed denote the commuting combinations
$Y_0=F_0\,,Y_1=F_{30}+"F_{13}"\,,Y_2=F_{10}+"F_{13}" \,;$
$\{Y_j\}$
also commute with the elements of the 
cyclic group $A$ as well as with the standard $sl(4)$
characters $\{\bc_{\bL} \}$. The  latter commute with any $w\in \tW$.
The  
fusions \fr, \fl{}, imply,
 postulating the expression for the character $\cc_{0130}$
as being given by the prescription in \mt,
 the following relations with
 coefficients $C_l\,, B_i^j\,,D$  in 
the group ring 
$\overline{\CR}[A]$, 
\eqn\rel{\eqalign{
&Y_0^2=4+Y_1+Y_2\,,\cr
&Y_j^2=C_j+ \sum_{i=0}^2\,B_i^j\, Y_i
\,,\quad j=1,2\,,\cr
&Y_0 (Y_1-Y_2)=D=\zg\, \bc_{\fw_1} - \zg^{3}\, \bc_{\fw_3}
\,.
}}
This suggests to consider linear combinations of the remaining 
five independent variables (e.g., $Y_0\,,Y_1\,,Y_2\,,Y_0 \,Y_1\,,
Y_1\,Y_2$), with coefficients
in  $\overline{\CR}[A]$. Requiring the validity of the multiplication
rule in \ws\ for any $\cc_x\in {\cal F}$ will impose
restrictions on these coefficients which might provide another
 route to the explicit construction of the characters.

\vskip 1cm
\noindent
{\tfont Acknowledgments}
\vskip 0.3cm

The first part of this note is a review of our joint papers \FGP\
with A.Ch. Ganchev, whom we thank for the collaboration.

P.F. acknowledges the support of the Italian Ministry of University,
 Scientific Research and Technology (MURST).
V.B.P. acknowledges the  support and hospitality of
INFN, Sezione di Trieste and ICTP, Trieste, the hospitality 
of the Arnold Sommerfeld Inst.\ f.\
Math.\ Phys.\ of TU Clausthal, as well as
partial support of the Bulgarian National Research Foundation
(contract $\Phi-643$).

\vskip 1cm

\noindent
\listrefs
\end